\documentclass[aps,prl,tightenlines,twocolumn,showpacs,preprintnumbers,amsmath,amssymb]{revtex4}

\usepackage{graphicx}
\usepackage{dcolumn}
\usepackage{bm}
\usepackage{epsfig}

\begin{document}

\title{ In-situ measurement of self-heating in intrinsic tunnelling spectroscopy. }

\author{V.M.Krasnov$^{1,2}$}
\author{M.Sandberg$^2$}
\author{I.Zogaj$^2$}

\affiliation {$^1$ Department of Physics, Stockholm University, Albanova University Center, SE-10691 Stockholm, Sweden\\
$^2$ Department of Microtechnology and Nanoscience, Chalmers
University of Technology, S-41296 G\"oteborg, Sweden}


\begin{abstract}

Using advanced sample engineering we performed simultaneous
measurement of interlayer tunnelling characteristics and in-situ
monitoring of temperature in Bi-2212 mesas. Together with a
systematic study of size-dependence of interlayer tunnelling this
allowed unambiguous discrimination between artifacts of
self-heating and gaps in electronic spectra of Bi-2212. Such a
confident spectroscopic information, which is not affected by
self-heating or surface deterioration, was obtained for the first
time for a High-$T_c$ superconductor. We also derived general
expressions and formulated main principles of self-heating valid
for a large variety of materials.

\end{abstract}


\maketitle

Surface spectroscopy of high temperature superconductors (HTSC)
experience considerable difficulties caused by rapid chemical
deterioration of the surface and a very short coherence length,
due to which surface and bulk properties can differ at a scale of
$\sim$ one atomic layer. Those problems are avoided in intrinsic
tunnelling spectroscopy (ITS), which utilizes intrinsic Josephson
junctions (IJJ's) naturally formed in highly anisotropic HTSC
\cite{Kleiner}. ITS has a superior resolution due to
superconductor-insulator-superconductor structure of IJJ's and
provides a unique opportunity to probe bulk electronic properties,
indispensable for understanding HTSC
\cite{Schlen,Kras_T,Suzuki,Kras_H}. Unfortunately, ITS is liable
to self-heating caused by poor thermal conductivity of HTSC
\cite{Heat_JAP,Gough,Heat_PhC,Suzuki_APL}. So far reports on
self-heating in ITS differ by orders of magnitude even in similar
short-pulse measurements \cite{Gough,Suzuki_APL}. In view of
unique abilities of ITS, lack of consensus between different
spectroscopic studies of HTSC, and recent controversy about the
role of self-heating in ITS \cite{Zavar,AYreply} it is important
to understand to what extent ITS is affected by self-heating.

Self-heating in superconductors is practically unstudied, even
though it is clear that it can be substantial due to inherently
poor thermal conductivity at low $T$. For example, it was
suspected for a long time that hysteresis in Current-Voltage
characteristics (IVC's) of overdamped Josephson junctions may
indicate a substantial self-heating. However, it is still unknown
to what extent self-heating affects spectroscopy of
superconductors even in more conventional cases of STM,
photoemission or IVC's of low-$T_c$ junctions. Self-heating is
also a growing problem for semiconducting devices often having a
mesa geometry \cite{Bayrak}.

Here we present a comprehensive study of self-heating in Bi-2212
mesa structures. Advanced sample engineering allowed simultaneous
measurement of IVC's and $T$ using a small portion of the mesa,
nano-patterned by Focused Ion Beam (FIB), as an in-situ
thermometer. Together with a systematic study of size-dependence
of ITS, this allowed unambiguous discrimination of self-heating
artifacts from gaps in electronic spectra of Bi-2212. Such a
confident spectroscopic information, not affected by self-heating
or surface deterioration, was obtained for the first time for a
HTSC. We also formulated main principles and derived general
expressions for self-heating in diffusive and ballistic cases.
Both the developed experimental technique and analytic results can
be used for studying self-heating in a large variety of materials.

To distinguish artifacts of self-heating we first analyze the
shape of IVC's, assuming that electronic Density of States (DoS)
is featureless, but the resistance exhibit strong dependence
$R(T)$~\cite{Zavar}:
\begin{equation}
V = IR(T_0 + \Delta T). \label{Eq.Zav}
\end{equation}
Here $T=T_0+\Delta T$ is the mesa temperature, which is higher
than the base temperature $T_0$ due to self-heating. We consider a
small circular mesa of radius $a$, containing $N$ IJJ's, on top of
a crystal of thickness $t$ with thermal conductivities
$\kappa_{ab,c}$ in the $ab-$plane and $c-$axis, respectively. Heat
flow has phononic and electronic channels. The latter is likely to
be diffusive and allows the exact solution~\cite{Heat_JAP}, which
for $a \ll 2 t \sqrt{\kappa_{ab}/\kappa_c}$ is
\begin{equation} \Delta T_{diff} = \frac{\pi}{4} \frac{q
a}{\sqrt{\kappa_{ab}\kappa_c}}, \label{Eq.Td2}
\end{equation}
where $q$ is a power density. However, phononic transport is
ballistic in the c-axis direction since phononic mean free path
$l_{ph} ~ \sim 1 \mu m$~\cite{Uher} is much larger than the height
of the mesa, containing few atomic layers. We estimate ballistic
contribution assuming that phonons first shoot to the depth $\sim
l_{ph}$ below the mesa and then spread via diffusion. In this case
$\Delta T$ can be estimated from Eq.(3) of Ref.\cite{Heat_JAP}
with $r=0$ and $z=l_{ph}$, multiplied by the fraction of heat,
$f$, carried by phonons. The total self-heating is the sum of
phononic and electronic contributions:

\begin{figure}
\noindent
\begin{minipage}{0.48\textwidth}
\epsfxsize=0.7\hsize \leftline{ \epsfbox{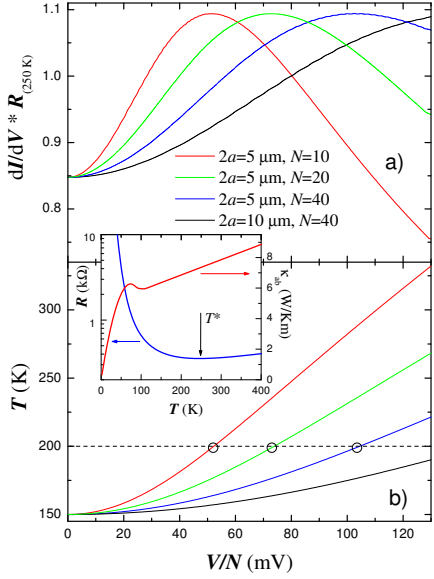}}
\caption{Simulations for the self-heating model: a) $dI/dV (V)$
and b) corresponding $T(V)$ curves for four mesas with different
geometry at $T_0 = 150 K$. A pronounced hump in $dI/dV$ is seen
for three largest mesas. Circles in Fig. b) indicate positions of
the humps. Irrespective of the sample geometry the hump occurs at
the same $T \simeq 200K$, eventhough the hump shape and voltage
are different. Inset shows $R(T)$ and $\kappa(T)$ dependencies
used in simulations. } \label{Fig.1}
\end{minipage}
\end{figure}
\begin{equation}
\Delta T \simeq \frac{q a }{2 \sqrt{\kappa_{ab}\kappa_c}}\left( f
\arctan \left[ \frac{a\sqrt{\kappa_c}}{l_{ph}\sqrt{\kappa_{ab}}}
\right] + (1-f)\frac{\pi}{2} \right) \label{Eq.Tbd}
\end{equation}
As expected, it coincides with Eq. (\ref{Eq.Td2}) for $l_{ph}
\rightarrow 0$. A large difference in anisotropies of electric
$\rho_c/\rho_{ab} \sim 10^5$ \cite{Watanabe,Kras_T} and thermal
$\kappa_{ab}/\kappa_c \sim 8$ \cite{Uher} conductivities implies
that $c-$axis heat transport in Bi-2212 is predominantly phononic,
$f \simeq 1$. Then we can write a simple estimation,
\begin{equation}
\Delta T_{ball} \simeq \frac{q a^2 }{4 \kappa_{ab} l_{ph}},
\label{Eq.Tb2}
\end{equation}
valid for $a < l_{ph}\sqrt{\kappa_{ab}/\kappa_c}$. Here we added a
factor $1/2$ since only $\sim 1/2$ of heat is going back to the
mesa.

To calculate IVC's we solved Eqs.(\ref{Eq.Zav},\ref{Eq.Td2}) with
$\kappa = \kappa(T_0 + \Delta T)$, using typical $R(T)$ and
$\kappa_{ab}(T)$ dependencies \cite{Uher}, shown in inset of Fig.
1, and $\kappa_{ab}/\kappa_c = 10$. Fig. 1 shows simulated $dI/dV
(V)$ and $T(V)$ curves for mesas with different geometry. Humps in
$dI/dV (V)$ appear for three largest mesas. The origin of humps is
easy to understand. At low $T_0$ heating leads to a decrease of
$R$. The ratio $V/I$ reaches minimum at the crossover temperature
$T^*$, see inset in Fig. 1. At higher bias, IVC's start to bend
backwards because $R(T)$ increases at $T> T^*$.

\begin{figure}
\noindent
\begin{minipage}{0.48\textwidth}
\epsfxsize=0.7\hsize \centerline{ \epsfbox{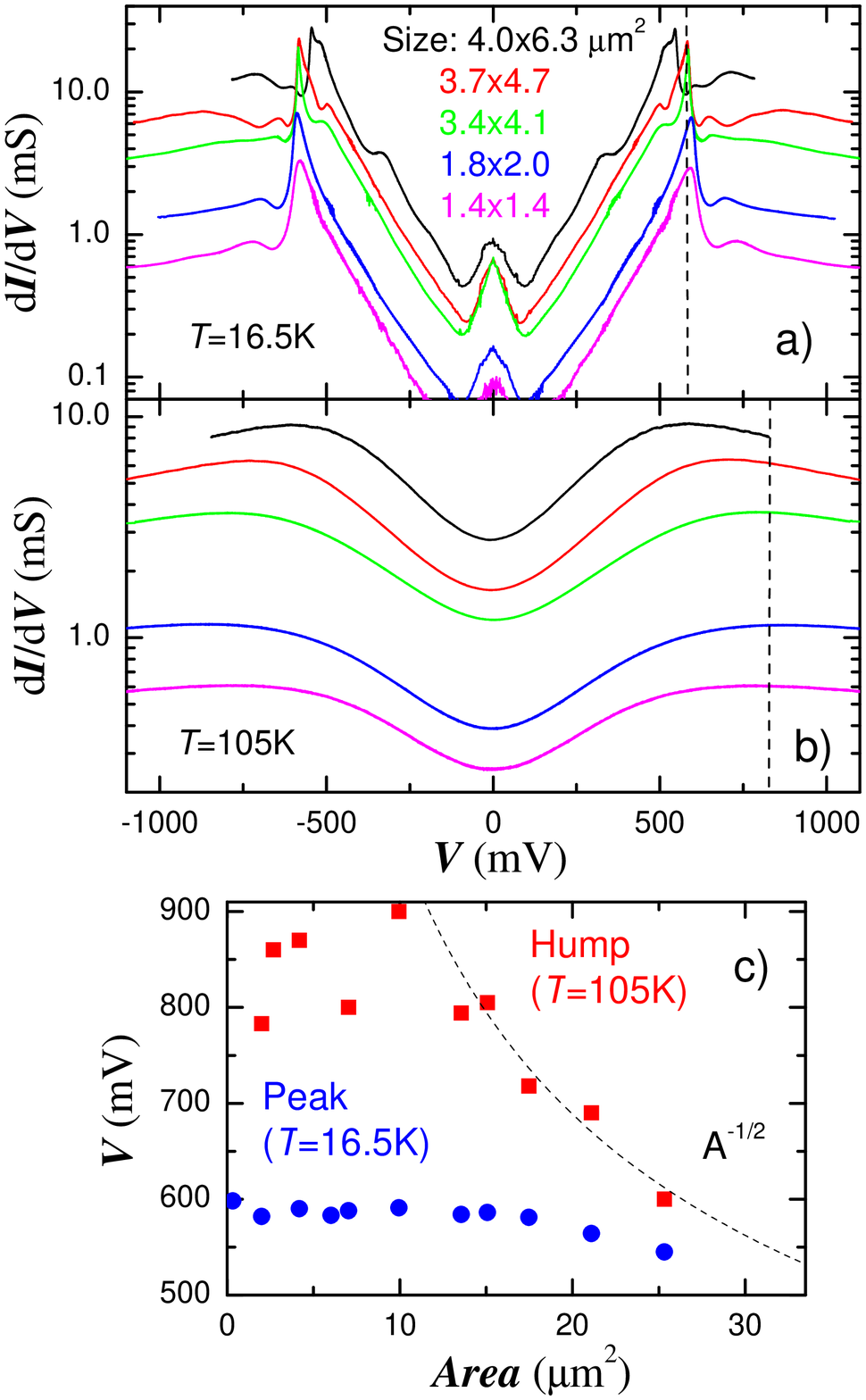}}
\caption{Experimental characteristics at a) $T_0 = 16.5 K$ and b)
$T_0 = 105 K$ for mesas with different area fabricated on the same
Bi-2212 single crystal. c) Size dependence of the peak voltage at
$T_0 = 16.5 K$ and the hump voltage at $T_0 = 105 K$. Dashed line
shows the fit for ballistic heating
.
}\label{Fig.2}
\end{minipage}
\end{figure}

Even though the shape of self-heating IVC's depends on details of
$R(T)$ and $\kappa(T)$, sample geometry and heat transport
mechanisms, artifacts of self-heating in all cases can be
understood from the following three principles:

1. $T-q$ curves are single valued. At a given $q$ mesas will be
heated to higher $T$ if started from higher $T_0$.

2. For mesas with the same $R(T)$, $\kappa(T)$ and $T_0$
self-heating features in IVC's occur at the same $T$ irrespective
of mesa geometry. This is clearly seen in Fig. 1 b).

3. Self-heating depends on the mesa size and the number of
junctions. From the second principle and
Eqs.(\ref{Eq.Td2},\ref{Eq.Tb2}) hump voltages in diffusive and
ballistic cases are:
\begin{eqnarray}
 v_{diff} = V/N = K_1/{\sqrt{a N}} \propto N^{-1/2}A^{-1/4},
\label{Eq.Vd}\\
v_{ball}= V/N = K_2/\sqrt{a^2 N} \propto N^{-1/2}A^{-1/2}.
\label{Eq.Vb}
\end{eqnarray}
Here $A$ is the mesa area and $K_{1,2} (T_0)$ are material
constants. The dependence Eq.(\ref{Eq.Vd}) can be clearly traced
from Fig. 1. This provides an unambiguous way to discriminate
electronic spectra from self-heating artifacts: \emph{Self-heating
depends on the sample geometry in contrast to the electronic
spectrum, which is a material property.}

\begin{figure}
\noindent
\begin{minipage}{0.48\textwidth}
\epsfxsize=0.9\hsize \centerline{ \epsfbox{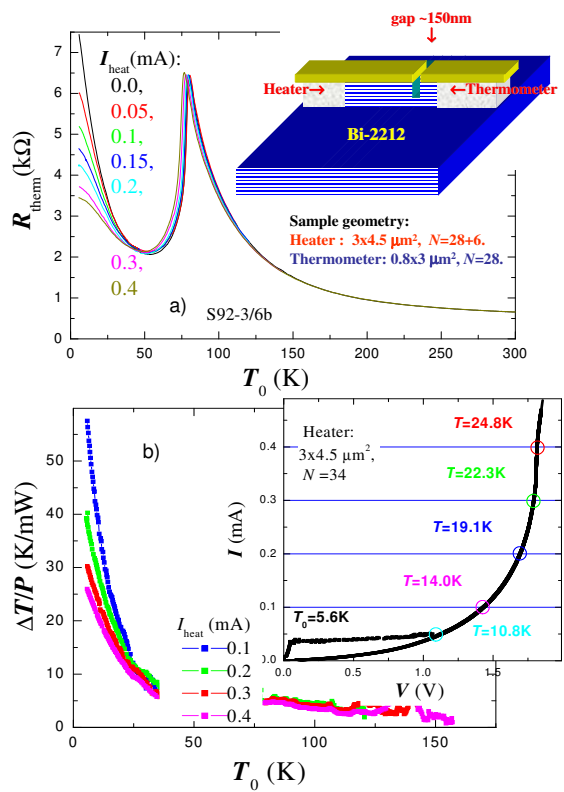} }
\caption{a) Thermometer mesa resistance as a function of the base
temperature for several currents through the heater mesa. Inset
shows the sample geometry. b) Normalized self-heating
characteristics for several heating currents. Inset shows the IVC
of the heater mesa at $T_0=5.6K$. The actual temperature along the
IVC is indicated.} \label{Fig.3}
\end{minipage}
\end{figure}

So far there was no systematic size-dependent ITS study. Since
self-heating depends on $R(T)$ and $\kappa(T)$, comparison should
be made for mesas at the same crystal. Fig. 2 shows $dI/dV (V)$
curves at $T_0 = 16.5 K$ and $105 K$ for mesas with different area
fabricated at the same optimally doped Bi-2212 single crystal,
$T_c \simeq 93 K$. All mesas contained $N=9$ IJJ's. Clearly
distinguishable a peak at $T_0 < T_c$ and a hump at $T_0 > T_c$
are seen, previously attributed to the superconducting gap (SG)
and the pseudo-gap (PG), respectively \cite{Kras_T,Suzuki}. Fig. 2
c) shows peak, $V_p$, and hump, $V_h$, voltages vs. mesa area. It
is seen that for small mesas $V_p$ is independent of area, but for
$A > 16 \mu m^2$, $V_p$ becomes slightly smaller due to
self-heating. This happens at $P(V_p)>1mW$. A stronger
size-dependence is seen for the hump. For $A> 10 \mu m^2$, $V_h$
is well described by the ballistic self-heating Eq.(\ref{Eq.Vb}),
as indicated by the dashed line. However, for $A < 10 \mu m^2$,
$V_h$ also shows a tendency for saturation. Here $P(V_h)<1.5 mW$.
Note that scattering of data for small mesas is caused by
flattening of the hump, see Fig. 2 b). It is not clear if the
residual flat hump represents the DoS, even though, STM
data~\cite{STM} and short pulse experiments \cite{Suzuki_APL},
which are less prone to self-heating, would imply so.
Nevertheless, for such a flat hump it is more instructive to
consider the characteristic slope of the V-shape $dI/dV(V)$ at
small bias, rather than $V_h$.

From Fig.2 it is seen that experimental ITS characteristics,
retain the same V-shape and slope below the peak and the hump,
irrespective of the mesa size. Such behavior is in stark contrast
to self-heating artifacts shown in Fig. 1 a), for which both the
slope and the voltage of the hump exhibited strong and correlated
size-dependence. Size-independence of the shape of ITS
characteristics together with saturation of peak and hump voltages
at $A \rightarrow 0$ implies that those indeed represent gaps in
DoS, rather than artifacts of self-heating. Such conclusion is
consistent with reported strong suppression of the peak by
magnetic field \cite{Kras_H}, which is hard to explain in terms of
self-heating, since $\kappa$ decreases with field \cite{Uher}.

The ultimate judgement about the extent of self-heating can be
made only by direct measurement of the mesa temperature. This
requires fabrication of an in-situ thermometer situated in an
intimate vicinity and having good thermal coupling to the mesa,
and allowing independent calibration. So far there were no
measurements which would satisfy all those requirements, even
though several attempts has been made \cite{Gough,AYreply}.

\begin{figure}
\noindent
\begin{minipage}{0.48\textwidth}
\epsfxsize=0.9\hsize \centerline{ \epsfbox{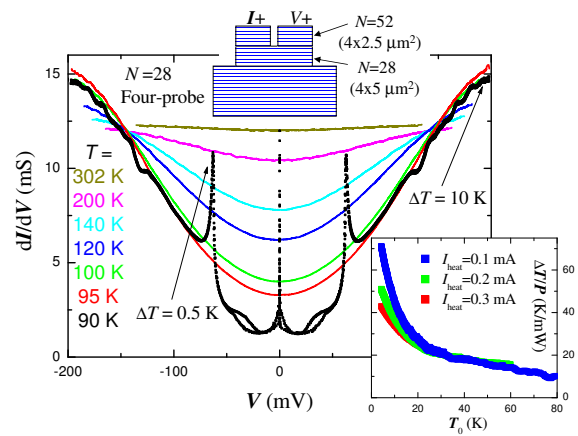} }
\caption{Four-probe ITS characteristics at $T_0$ slightly below
and above $T_c\simeq 93 K$. Sample geometry is shown in the top
inset. Bottom inset shows in-situ measured self-heating $\Delta
T/P$. } \label{Fig.4}
\end{minipage}
\end{figure}

To facilitate such measurements we developed a new method for
fabrication of nano-scale multi-terminal arrays of IJJ's: first a
mesa was made on top of a Bi-2212 crystal by self-alignment
cross-bar photolithography. Then a small portion of the initial
mesa was cut using FIB. The bias current was applied through the
larger mesa (the heater) and the $R(T)$ dependence of the nearby
smaller mesa (the thermometer) was used for in-situ measurement of
$T$ \cite{Gough}. Typically we stopped the FIB cut within the
height of the initial mesa so that both the heater and the
thermometer mesas were sitting on a common pedestal - the bottom
part of the initial mesa. The sketch of the sample geometry is
shown in inset of Fig. 3 a). Nano-scale separation between the
heater and the thermometer mesas and small $\nabla T$ in the
pedestal \cite{Heat_JAP} facilitated accurate measurement of
self-heating.

Fig. 3 a) shows $R(T_0)$ of the thermometer for a set of
dc-currents through the heater mesa, $I_{heat}$, for an underdoped
Bi-2212 crystal, $T_c \simeq 81 K$. It is seen that $R$ decreases
with $I_{heat}$ as a result of heating. The $\Delta T$ can be
obtained by subtracting $T_0$ vs $R$ curves at certain $I_{heat}$
from that without heating, $I_{heat}=0$. Corresponding
self-heating characteristics, normalized by the dissipation power
are shown in Fig. 3 b). It is seen that at low $T_0$, $\Delta T /
P$ decreases with $I_{heat}$. This is due to a strong $\kappa (T)$
dependence at low $T$, see inset in Fig. 1. At higher $T_0 > 40-50
K$, $\Delta T / P$ curves for different $I_{heat}$ approach each
other, indicating flattening of $\kappa (T)$ dependence.

Inset in Fig. 3 b) shows the IVC of the heater mesa at $T_0 = 5.6
K$. The actual $T$ along the IVC is indicated. It is seen that at
the end of sum-gap knee (above the peak) the mesa is heated by
$19.2 K$, i.e. considerably smaller than $T_c$. The peak voltage
is not affected by such heating \cite{Heat_PhC} since the SG has a
flat dependence in this $T-$range~\cite{Kras_T}.

Sample geometry used in this work is also interesting because it
facilitates true four-probe measurements of IJJ's in the pedestal,
removing questions about the influence of contact resistance
between the mesa and the electrode and improving ITS resolution.
Fig. 4 shows four-probe characteristics from $90 K$ to $302 K$.
Here we can clearly observe the peak just few $K$ below $T_c\simeq
93K$ as well as some, presumably strong-coupling phonon features
above the SG, which disappear at $T_c$. Inset in Fig. 4 shows
in-situ measured self-heating $\Delta T/P$ for this mesa, which is
similar to that in Fig. 3 b) and all other measured samples. In
Fig. 4 it is indicated that self-heating at the peak at $T_0 = 90
K$ is small $\Delta T \sim 0.5 K$ even despite a large number of
IJJ's, $N=90$ in total. Thus, influence of self-heating on the
peak vanishes at $T_0 \rightarrow T_c$ because both $\Delta T/P$
decreases with $T$ and the power at the peak vanishes together
with the SG at $T_0 \rightarrow T_c$ \cite{Kras_T}. The realistic
simulation of how self-heating affects $T-$dependence of the SG
can be found in Ref.\cite{Heat_PhC}.

From Fig. 4 it is seen that $dI/dV (V)$ are V-shaped at $T>T_c$
with a pronounced dip at $V=0$. Whether or not the dip is a
self-heating artifact can be understood from in-situ $T-$
measurements, shown in inset. In Fig. 4 it is indicated that at
$T_0 = 90 K$ the mesa is heated to $\sim 100 K$ at $V = 200 mV$.
On the other hand, at $T_0 = 100K$ the self-heating free
$dI/dV(V=0)\simeq 4.0 mS$ is considerably smaller than
$dI/dV(V=200mV)\simeq 14.8 mS$. Therefore, self-heating is
insufficient to explain such a large increase of conductance from
$V=0$ to $200 mV$ and the V-shape characteristics must be
attributed to the PG in DoS.

Now we can reanalyze one of the most important results of previous
ITS studies - evidence for coexistence of the SG and the
PG~\cite{Kras_T,Suzuki,Kras_H}, which favors different origins of
the two gaps. When the peak and the hump are observed
simultaneously, the hump occurs at larger $V$ and, therefore, at
higher $T$ than the peak. From Fig. 2 it is seen that for large
mesas the hump is affected by self-heating and is exaggerated due
to progressive back-bending of $dI/dV(V)$ at large $V$. In most of
previous ITS experiments dealing with larger mesas, humps were
certainly affected by self-heating. This is particularly true for
Bi-2201 compound~\cite{AYreply}, which has a similar $\Delta T/P$
as Bi-2212, but considerably lower $T_c$ and, thus, more prone to
self-heating. So, does the PG exists at $T < T_c$? The answer can
be obtained from Fig. 4. It is seen that the $dI/dV (V)$ curve at
$T_0 =90 K$ acquires characteristic V-shape right after the peak,
where self-heating is still small. Thus, the PG does coexist with
superconductivity at $T < T_c$. Such conclusion is also supported
by observation of V-shape PG characteristics at $T<T_c$ in
magnetic fields $H>H_{c2}$ \cite{Kras_H} and in the vortex core
\cite{STM}.

In summary, we performed a comprehensive analysis of self-heating
in Bi-2212 mesas, which allowed unambiguous discrimination of gaps
in electronic spectra from artifacts of self-heating. This was
achieved via systematic size-dependent study and in-situ
measurement of self-heating using a nano-patterned part of the
mesa as the in-situ thermometer. We observed that for small mesas
both shapes and voltages of the peak and the hump (or rather
V-shaped suppression of conductance at $V=0$) are independent of
the mesa size and appear at $T$ considerably smaller than $T_c$
and $T^*$, respectively. Therefore, they are not due to
self-heating \cite{Zavar}, but represent the superconducting gap
and the normal state pseudo-gap in electronic DoS, respectively.
Such a confident spectroscopic information for a HTSC not affected
by self-heating, or surface deterioration was obtained here for
the first time. On the other hand, ITS characteristics of larger
mesas can be strongly affected by self-heating. The reported
threshold mesa size and dissipation power, at which self-heating
becomes insignificant, as well as the measured values of
self-heating $\Delta T/P$ are typical for our mesas, but not
universal. They depend on the sample geometry, materials and
experimental setup. Thus, it is important to carefully design
samples for ITS: decrease mesa sizes, number of junctions, avoid
suspended structures and, in particular, to employ the top heat
spreading layer, which is the most important heat sinking channel
in mesas~\cite{Heat_JAP,Bayrak}.

Finally we want to note that derived expressions and formulated
principles of self-heating are general and valid for any material.
Similarly, the developed experimental method, in which a small
portion of the sample is used as the in-situ detector, can be used
for analysis of self-heating in a large variety of materials. The
only requirement is that the resistance of material or the contact
between electrode and material should have considerable
$T-$dependence, while layered structure and HTSC is not essential.
For example, our method can be directly applied for studying
self-heating in small mesa-like semiconducting transistors
\cite{Bayrak}.

Financial support from the Swedish Research Council is gratefully
acknowledged.

\section{Erratum}
The units of the vertical axis in Fig. 3a) should read
$10k\Omega$, not $k\Omega$. This does not affect any part of the
manuscript because only relative shapes of the curves were used
for determining shifts in temperature.


\begin{references}

\bibitem{Kleiner} R. Kleiner and
P.~M\"{u}ller, {\em Phys. Rev. B} {\bf 49}, 1327 (1994);
V.M.Krasnov, et al., {\em ibid.} {\bf 59}, 8463 (1999); H.Wang, et
al., {\em Phys. Rev. Lett.} {\bf 87}, 107002 (2001);

\bibitem{Schlen} K.~Schlenga et al., {\em Phys. Rev. B} {\bf 57}, 14518
(1998)

\bibitem{Kras_T} V.M.~Krasnov et al., {\em Phys.Rev.Lett.} {\bf 84},
5860 (2000); {\em Phys.Rev.B} {\bf 65}, 140504(R) (2002)

\bibitem{Suzuki} M.Suzuki, et.al., {\em Phys.Rev.Lett.} {\bf 82},
5361 (1999); {\em ibid.} {\bf 85}, 4787 (2000)

\bibitem{Kras_H} V.M.~Krasnov et al., {\em Phys.Rev.Lett.} {\bf 86},
2657 (2001)

\bibitem{Heat_JAP} V.M.Krasnov et al., {\em J.Appl.Phys.} {\bf 89}, 5578
(2001); {\em ibid.} {\bf 93}, 1329 (2003);

\bibitem{Gough} P.J.Thomas et al., {\em Physica C} {\bf 341-348},
1547 (2000); J.C.Fenton, et al., {\em Appl.Phys.Lett} {\bf 80},
2535 (2002)

\bibitem{Heat_PhC} V.M.~Krasnov, {\em Physica C} {\bf 372-376},
103 (2002)

\bibitem{Suzuki_APL}  K.~Anagawa et al., {\em Appl.Phys.Lett.} {\bf 83}, 2381 (2003)

\bibitem{Zavar} V.N.~Zavaritsky, {\em Phys.Rev.Lett.} {\bf 92},
259701 (2004); {\em Physica C} {\bf 404}, 440 (2004)


\bibitem{AYreply} A.Yurgens et al., {\em Phys.Rev.Lett.} {\bf 92},
259702 (2004); {\em cond-mat}/{\bf 0309131}

\bibitem{Bayrak} B.Bayraktaroglu et al., {\em IEEE Trans.El.Dev.Lett.}
{\bf 14}, 493 (1993)

\bibitem{Watanabe} T. Watanabe et al., {\em Phys.Rev.Lett.} {\bf
79}, 2113 (1997)

\bibitem{Uher} C.~Uher, in {\it Physical Properties of High Temperature Superconductors III}, ed. D.M.~Ginsberg (World Sci. Publ. Co., Singapore 1992) Vol. 3, p.
159;

Y.Ando, et.al, Phys.Rev.B 62 (2000) 626


\bibitem{STM} Ch.Renner et al., {\em Phys. Rev. Lett} {\bf 80},
149 (1998); B.W.Hoogenboom et al., {\em ibid.} {\bf 87}, 267001
(2001)

\end{references}
\end{document}